\documentclass[lettersize,journal]{IEEEtran}
\usepackage{amsmath,amsfonts}
\usepackage{algorithmic}
\usepackage{algorithm}
\usepackage{array}
\usepackage[caption=false,font=normalsize,labelfont=sf,textfont=sf]{subfig}
\usepackage[colorlinks,citecolor=blue,linkcolor=blue,urlcolor=blue]{hyperref}
\usepackage{textcomp}
\usepackage{stfloats}
\usepackage{url}
\usepackage{verbatim}
\usepackage{graphicx}
\usepackage{cite}
\usepackage{tabularx}
\usepackage{color}
\usepackage{booktabs}
\usepackage{fontspec}
\usepackage{newtxtext}
\usepackage{balance}

\newcolumntype{C}[1]{>{\centering\arraybackslash}m{#1}}

\newcolumntype{L}{>{\raggedright\arraybackslash}X}

\begin{document}

\title{Fluid Antennas Meet Rate-Splitting Multiple Access: \\A New Path Forward for 6G Networks}

\author{Jinyuan~Liu,
        Yong~Liang~Guan,~\IEEEmembership{Senior~Member,~IEEE,}
        Hong~Niu, Qian~Zhang,    \\
        M\'{e}rouane~Debbah,~\IEEEmembership{Fellow, IEEE,}
        Hyundong~Shin,~\IEEEmembership{Fellow, IEEE,}
        and~Bruno~Clerckx,~\IEEEmembership{Fellow,~IEEE}
\thanks{Jinyuan Liu, Hong Niu, and Yong Liang Guan are with the School of Electrical and Electronic Engineering, Nanyang Technological University, Singapore 639798 (e-mail:
jinyuan001@e.ntu.edu.sg; hong.niu@ntu.edu.sg; eylguan@ntu.edu.sg).}
\thanks{Qian Zhang is with the School of Information Science and Engineering, Shandong University, Qingdao, China. (e-mail: qianzhang2021@mail.sdu.edu.cn).}
\thanks{M. Debbah is with the Research Institute for Digital Future, Khalifa University, 127788 Abu Dhabi, UAE (email: merouane.debbah@ku.ac.ae).}
\thanks{H. Shin is with the Department of Electronics and Information Convergence Engineering, Kyung Hee University, Yongin-si, Gyeonggi-do 17104, Republic of Korea (e-mail: hshin@khu.ac.kr).}
\thanks{Bruno Clerckx is with the Department of Electrical and Electronic Engineering, Imperial College London, London SW7 2AZ, U.K. (e-mail: b.clerckx@imperial.ac.uk).}
}



\maketitle

\begin{abstract}
Future sixth-generation (6G) networks require high spectral efficiency (SE), massive connectivity, and stringent reliability under imperfect channel state information at the transmitter. Rate-splitting multiple access (RSMA) addresses part of this challenge by flexibly managing interference through common and private message streams, while fluid antenna systems (FAS) offer low-cost spatial diversity by dynamically reconfiguring antenna positions within a compact aperture. In this paper, we first classify FAS-enabled multiple access systems from the perspectives of FAS deployment, objectives, and antenna configuration, along with some comparisons with benchmark schemes, thereby exhibiting the inherent efficiency of FAS-RSMA. Moreover, we reveal the mutually enhancing mechanism between FAS and RSMA: FAS strengthens the weakest effective link and improves the beamforming design in RSMA, whereas RSMA turns FAS-induced spatial diversity into robust interference management under diverse channel conditions. In addition, we identify representative 6G scenarios and highlight major research challenges in joint beamforming-antenna position design, channel estimation, and hardware design. Furthermore, case studies quantify the gains of FAS-RSMA over the fixed-position antenna (FPA) system with RSMA and NOMA baselines, which validates that FAS-RSMA is a strong candidate for interference-limited access in 6G systems.
\end{abstract}


\section{Introduction}
Sixth-generation (6G) networks aim to support higher spectral efficiency (SE), user throughput, and device connectivity under stringent latency, reliability, and quality-of-service (QoS) requirements, even with imperfect channel state information at the transmitter (CSIT) in dense deployments.

Compared to conventional orthogonal multiple access schemes, rate-splitting multiple access (RSMA) has recently gained attention as a compelling physical-layer strategy thanks to its SE and operational flexibility \cite{ref1}. The roots of rate-splitting trace back to the two-user interference channel, with subsequent developments adapting the idea to downlink multi-user multi-antenna transmission \cite{ref2}. RSMA splits each user's message into common and private parts, encodes them into common and private streams respectively, and receivers use successive interference cancellation (SIC) to first decode and cancel the common stream before decoding the private stream, thereby realizing a hybrid interference management strategy where part of the interference is decoded and part of the interference is treated as noise \cite{ref3}. Owing to this flexibility, RSMA has been investigated across a wide range of scenarios \cite{ref1}, which consistently demonstrates its reliability and flexibility compared with other multiple access schemes.

RSMA has been widely recognized for its superior ability to manage interference and enhance SE in wireless networks, and is typically implemented on multi-antenna transceivers that follow the conventional one-RF-chain-per-antenna architecture. As 6G evolves towards larger arrays, denser deployments, and strictly power-constrained terminals, scaling up this architecture incurs prohibitive hardware costs and energy consumption \cite{ref4}. This motivates the exploration of complementary antenna architectures that can expose spatial flexibility under a limited RF-chain budget. In this context, fluid antenna systems (FAS)\footnote{{To avoid ambiguity, we note that FAS includes any software-controllable structure, including liquid-based antennas, pixel-based antennas, and metasurfaces, capable of dynamically reconfiguring its position, shape, or other radiation characteristics \cite{ref4}. Thus, the term ``fluid'' highlights this remarkable spatial agility rather than strictly referring to a physical liquid medium.}} have been explored to enhance reliability in large-scale MIMO for 6G networks \cite{ref5}. FAS employs liquid-mediated radiators or reconfigurable pixel elements whose effective radiating position can be relocated within a constrained aperture. By steering the active location, the terminal exploits additional spatial degrees of freedom (DoFs) and realizes substantial selection-diversity gains even when the device footprint is limited. {The achievable reconfiguration (switching) latency is highly dependent on the underlying hardware architecture. In general, liquid-mediated or mechanically movable implementations may incur relatively slower switching response times, typically in the millisecond-to-second range. Conversely, pixel-based electronically reconfigurable structures can support significantly faster switching, ranging from nanoseconds to microseconds, thereby enabling highly agile spatial adaptation when tracking fast-varying channels.}

The complementary strengths of RSMA and FAS naturally motivate their joint design \cite{ref6,ref7,ref8,ref9,ref10,ref11,ref12,ref13,ref14,ref15}. RSMA provides robust interference management and rate adaptation through message splitting into common and private streams \cite{ref3}, and can be implemented on conventional fully-digital or hybrid beamforming arrays \cite{ref6}. In large-array or cost- and power-constrained settings, however, it is also attractive to consider alternative antenna architectures that expose spatial flexibility under a very limited RF-chain budget. In this context, FAS offers adaptive antenna position reconfiguration within a compact aperture\footnote{{Reconfigurable intelligent surface (RIS) can complement FAS-RSMA by shaping the incident field and creating additional controllable paths, which can partially compensate for a limited FAS travel range by enriching the effective channel variation across candidate ports.}} \cite{ref7,ref8}. In an FAS-RSMA architecture, dynamic antenna position reconfiguration can enhance user signal strength compared with fixed-position antenna (FPA) systems under a limited RF-chain budget, thereby improving data stream reliability and easing the SIC process \cite{ref9,ref14,ref15}. Conversely, RSMA enhances FAS operation by flexibly managing interference and power allocation under imperfect CSIT and diverse user channels, translating the spatial diversity created by antenna positions mobility into consistent gains in reliability and throughput \cite{ref10,ref11,ref13}. Taken together, these benefits position FAS-RSMA as a promising and complementary approach for 6G network design.

The core role of this paper is to build a clear picture of \emph{what} advantages FAS-RSMA can offer and \emph{how} these advantages can be harnessed in future 6G systems. We synthesize existing results into a simple taxonomy and compare FAS-RSMA with FPA-RSMA and FAS/FPA-NOMA in terms of interference management, robustness, and diversity. Building on this, we outline where FAS-RSMA is most likely to be used in 6G and highlight the key design questions that need to be answered for these deployments. In this way, the paper serves as a concise guide for researchers and practitioners seeking to understand and leverage FAS-RSMA as a promising multiple-access option for 6G.

\begin{figure}[ht]
    \setlength{\belowcaptionskip}{-0.9cm}
    \centering
    \includegraphics[width=0.5\textwidth]{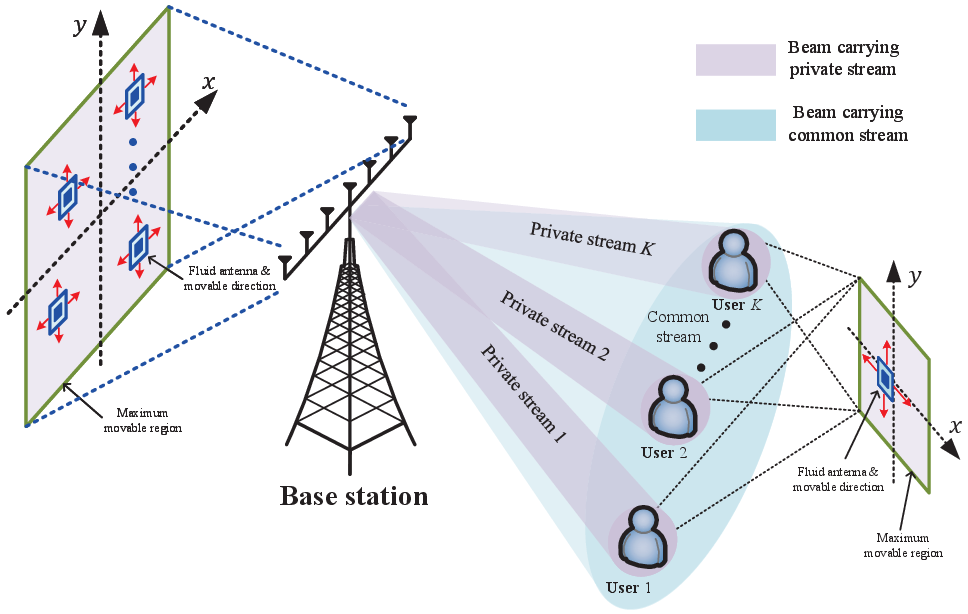}
    \caption{The multi-user downlink FAS-RSMA system framework.}
    \label{fig:System model}
\end{figure}

\section{Classification of FAS-Enabled Multiple Access Systems}
\begin{table*}[!t]
    \centering
    \caption{Overview of FAS-Enabled Multiple Access Systems}
    \label{table1}
    \renewcommand{\arraystretch}{1.3} 
    \setlength{\tabcolsep}{4pt}
    
    \begin{tabularx}{\textwidth}{|C{0.05\textwidth}|C{0.1\textwidth}|C{0.1\textwidth}|C{0.1\textwidth}|L|}
        \hline
        \textbf{Refs.} & \textbf{FAS Deploy.} & \textbf{Objectives} & \textbf{Ant. Configs.} & \textbf{Key features} \\
        \hline
 
        \cite{ref6} & F4 
              & O2, O3
              & A2
              & Demonstrates RSMA in multi-antenna downlink can simultaneously improve SE and EE, providing a non-FAS RSMA baseline that highlights the spatial flexibility FAS can offer. \\
        \hline
        \cite{ref7} & F3 
              & O1, O2
              & A3
              & Provides an information-theoretic characterization of MIMO-FAS, which establishes that FAS can offer spatial flexibility that is highly synergistic with RSMA’s interference management. \\
        \hline
        \cite{ref8} & F2
              & O2
              & A1
              & Explores user fairness in FAS-NOMA, showing that antenna positions reconfiguration improves OP balance, which underlines FAS as a fairness enabler that can be further leveraged in FAS-RSMA designs. \\
        \hline
        \cite{ref9} & F2
              & O2
              & A1
              & Studies a SISO downlink with FAS-RSMA and derives OP expressions, showing that antenna positions reconfiguration in FAS-RSMA improves reliability over fixed-antenna RSMA and NOMA. \\
        \hline
        \cite{ref10} & F2
              & O1
              & A2
              & Incorporates FAS-RSMA into the integrated sensing and communications (ISAC) framework, and jointly optimizes antenna positions selection and precoding under channel uncertainty to enhance both secure communication and sensing robustness.\\
        \hline
        \cite{ref11} & F1
              & O1
              & A2
              & Proposes FAS-RSMA architecture for non-terrestrial networks, showing that joint adaptation of satellite beams and FAS antenna positions sustains high throughput and fairness under highly dynamic channels. \\
        \hline
        \cite{ref12} & F3
              & O1, O3
              & A2
              & Considers a wireless-powered NOMA network with fluid/movable antennas, and optimizes antenna positions to illustrates how fluid arrays can improve throughput and EE in non-orthogonal access. \\
        \hline
        \cite{ref13} & F1
              & O1
              & A2
              & Investigates RSMA with movable antennas for short-packet ultra-reliable low-latency communication (URLLC). Results indicate that combining RSMA with movable/FAS antennas provides additional spatial DoFs. \\
        \hline
        \cite{ref14} & F2
              & O2
              & A1
              & Analyzes an unmanned aerial vehicle (UAV)-relay FAS-RSMA system, which shows that properly chosen UAV positions and FAS port configurations improve RSMA link robustness, especially in coverage-limited scenarios. \\
        \hline 
        \cite{ref15} & F2
              & O2
              & A1
              & Develops a correlation-aware analytical framework for FAS-RSMA and shows that FAS-RSMA can improve reliability and interference-management ability compared with FPA RSMA/NOMA. \\
        \hline        
    \end{tabularx}
\end{table*}
An illustrative FAS-RSMA system model is depicted in Fig. \ref{fig:System model}, where a multi-antenna base station (BS) equipped with FAS serves multiple users, each also endowed with FAS capabilities. {To ensure clarity, we define the key terms used throughout this paper. A candidate port refers to a predefined location within the compact aperture of the FAS where a signal can be radiated or received. Activation denotes the process of connecting a selected candidate port to the RF chain. Consequently, the considered port reconfiguration involves switching the RF connection among discrete candidate ports, bypassing the need for continuous physical antenna movement.} In this section, we review the literature on FAS-enabled multiple access with a particular emphasis on how these works motivate and position FAS-RSMA. The papers in Table~\ref{table1} are organized along three axes: FAS deployment, system objectives, and antenna configurations, which together reveal how FAS and RSMA can be jointly exploited. Works on FAS-assisted NOMA and on FPA-RSMA serve as baselines.
\subsection{FAS Deployment}
FAS deployment determines the control over the effective channels at the BS and users. These configurations fall into four categories:
\begin{itemize}
    \item \textbf{F1}: FAS-enabled BS. A BS equipped with FAS aperture exposes many candidate ports and activates only a subset at any time, based on the radio environment and service targets. Port reconfiguration is jointly designed with the multi-user precoder and the common-private rate split.

    \item \textbf{F2}: FAS-enabled users. At the users, FAS offers multiple candidate ports, from which the device activates those that best match its instantaneous channel and interference conditions. 

    \item \textbf{F3}: FAS-enabled BS and users. Both the BS and the users are equipped with FAS, enabling joint optimization of transmit and receive antenna positions. This bidirectional reconfigurability provides the richest spatial DoFs for FAS-RSMA.

    \item \textbf{F4}: FPA-enabled BS and users. Both the BS and the users rely on FPA with no port reconfiguration. This deployment serves as a FPA baseline to quantify the additional gains brought by FAS-enabled RSMA architectures.
\end{itemize}

\subsection{Objectives}
Depending on the system configuration and design focus, the objectives can be grouped into three main categories:
\begin{itemize}
    \item \textbf{O1}: Achievable rate. Rate-related metrics such as instantaneous achievable rate, sum-rate, and secrecy rate are used to quantify SE and physical-layer security. In FAS-RSMA, these metrics capture how well port reconfiguration and rate-splitting jointly exploit spatial DoFs to boost throughput or secure rate.

    \item \textbf{O2}: Outage probability (OP). OP measures the probability that the instantaneous rate or signal-to-interference-plus-noise ratio (SINR) falls below a target threshold, thus focusing on reliability rather than SE. For FAS-RSMA, OP is especially relevant as port reconfiguration and multi-user interference jointly determine whether common and private streams can be decoded without interruption.

    \item \textbf{O3}: Energy efficiency (EE). EE-oriented designs seek to maximize the number of reliably delivered bits per unit of consumed energy, accounting for both transmit and circuit power, including the cost of activating additional antenna positions and RF chains. FAS-RSMA fits naturally into this framework by selectively activating antenna positions while maintaining robust interference management.
\end{itemize}

\subsection{Antenna Configurations}
The BS and user antenna configurations govern FAS-RSMA spatial DoFs and receiver complexity. We categorize them into three types:
\begin{itemize}
    \item  \textbf{A1}: Single-input single-output (SISO). The BS uses a single transmit antenna, while each user terminal is typically endowed with FAS and performs port reconfiguration to activate the most favorable port. This configuration is particularly useful to quantify how much FAS-RSMA can gain from receiver-side port diversity even without multiple transmit antennas.

    \item \textbf{A2}: Multiple-input single-output (MISO). The BS is equipped with multiple transmit antennas and exploits beamforming to manage inter-user interference, while FAS can be placed at the BS, the UEs, or both. Such a configuration provides a practical basis for FAS-RSMA designs in downlink cellular systems, as it combines conventional multi-antenna precoding with additional spatial flexibility brought by port reconfiguration.

    \item \textbf{A3}: Multiple-input multiple-output (MIMO). Both the BS and the users are equipped with multiple antennas, enabling joint transmit beamforming and receive combining. When integrated with FAS, such a configuration provides the largest potential gains for FAS-RSMA, but also requires the highest design and hardware complexity.
\end{itemize}
Overall, this taxonomy highlights a consistent trend: combining FAS with flexible multiple access leverages port reconfiguration to enhance effective link quality, improve rate balance, and mitigate multi-user interference. RSMA aligns well with this framework, since its common-private structure allows FAS to strengthen the effective channels while RSMA controls interference and rate allocation.

\begin{table*}[!t]
    \centering
    \caption{Comparison of FAS-RSMA and Benchmark Multiple Access Schemes}
    \label{tab:FAS_RSMA_comparison}
    \renewcommand{\arraystretch}{1.3} 
    \setlength{\tabcolsep}{4pt}
    
    \begin{tabularx}{\textwidth}{|C{0.09\textwidth}|C{0.09\textwidth}|C{0.16\textwidth}|C{0.09\textwidth}|C{0.09\textwidth}|L|}
        \hline
        \textbf{Schemes} 
        & \textbf{Spatial Adaptation} 
        & \textbf{Interference Management Ability} 
        & \textbf{Robustness} 
        & \textbf{Diversity Gain} 
        & \textbf{Distinctive Features} \\
        \hline
        FAS-RSMA 
        & Dynamic 
        & Very high 
        & Very high 
        & High 
        & Jointly exploits FA port reconfiguration and rate-splitting to shape both the effective channels and multi-user interference. \\
        \hline
        FPA-RSMA 
        & None 
        & High 
        & High 
        & Moderate 
        & Adopts RSMA on fixed antenna arrays to serve as a non-FAS baseline lacking port reconfiguration capability. \\
        \hline
        FAS-NOMA 
        & Dynamic 
        & Moderate 
        & Moderate 
        & Moderate 
        & Uses port diversity mainly to enlarge channel disparities and support NOMA pairing. \\
        \hline
        FPA-NOMA 
        & None  
        & Low
        & Low 
        & Low
        & Employs conventional NOMA without FA or rate-splitting to highlight the performance limitations that FAS-RSMA aims to overcome. \\
        \hline
    \end{tabularx}
\end{table*}

\section{Rationale and Advantages of FAS-RSMA}
FAS-RSMA sits at the intersection of two complementary capabilities: hardware-level spatial adaptation via FA and signal-level interference management via rate-splitting. Table~\ref{tab:FAS_RSMA_comparison} contrasts FAS-RSMA with FPA-RSMA, FAS-NOMA, and FPA-NOMA along four dimensions, including spatial adaptation, interference management, robustness, and diversity gain.

In the remainder of this section, we clarify how FAS enhances RSMA and, conversely, how RSMA benefits FAS, and we relate these qualitative insights to the attributes summarized in Table~\ref{tab:FAS_RSMA_comparison}. Considering these two complementary perspectives helps to build a clearer picture of the performance improvements and design opportunities offered by jointly integrating FAS and RSMA.
\subsection{How FAS Enhances RSMA}
From the RSMA perspective, FAS can bring enriching spatial DoFs through dynamic port reconfiguration, which leads to the following benefits.
\subsubsection{\textbf{Mitigating channel-imposed constraints}}
In practical RSMA deployments, the rate of the common stream must be chosen such that all scheduled users can decode it successfully \cite{ref15}. This fundamental broadcast constraint effectively couples the common-stream rate to the user experiencing the weakest instantaneous channel. Introducing FAS helps to alleviate this physical propagation challenge: port-reconfiguration enhances the effective channel gain of users in deep fades, increases the decoding margin for the common stream, and thereby relaxes the constraint imposed by the weakest user \cite{ref15}. In terms of Table~\ref{tab:FAS_RSMA_comparison}, this evolution from static antenna positions to dynamic spatial adaptation directly contributes to the higher robustness and diversity gain of FAS-RSMA compared with FPA-RSMA.

\subsubsection{\textbf{Improving the beamforming design}}
Equipping the BS with FAS introduces reconfigurable spatial DoFs, enabling the array geometry and active port set to be adapted to the instantaneous propagation environment \cite{ref11}. By strengthening the effective channel gains and reducing inter-user channel correlation, BS-side FAS facilitates more effective data stream beamforming design, thereby enhancing the overall beamforming gain. As a result, users experience higher SINR, which translates into improved achievable rates, reduced OP, and enhanced fairness \cite{ref13}. Compared with RSMA operating on FPA, FAS therefore provides a more flexible and hardware-efficient means of exploiting spatial DoFs to boost system-level performance, which is reflected in the \emph{very high} diversity gain and \emph{very strong} interference management ability attributed to FAS-RSMA in Table~\ref{tab:FAS_RSMA_comparison}.

\subsubsection{\textbf{Leveraging mobility in realistic channels}}
Beyond static deployments, implementations of FAS at the BS or users enable continuous adjustment of antenna positions in response to channel variations. When integrated with RSMA, the joint optimization of port reconfiguration and transmit beamformers can maximize the sum rate or URLLC-oriented performance metrics, thereby achieving higher data rates and reliability with fewer RF components than conventional fixed arrays \cite{ref13}. These findings indicate that FAS endows RSMA with an additional spatial adaptation DoFs, which is particularly beneficial in rapidly time-varying or strongly correlated propagation environments, such as UAV-relay systems and non-terrestrial networks \cite{ref14}. This additional adaptation explains why FAS-RSMA achieves stronger reliability and diversity gain than FPA-RSMA.

\subsection{How RSMA Enhances FAS}
From the viewpoint of FAS, RSMA offers a flexible, interference-aware transmission framework that can more effectively manage multi-user interference, facilitate fairness, guarantee reliability, and support multi-objective designs such as secure communication and sensing-aware transmission. These advantages can be summarized as follows.

\subsubsection{\textbf{Improving interference management ability}}
While FAS enriches the physical layer with additional spatial DoFs, it still requires an appropriate multi-user transmission strategy to fully exploit these capabilities and handle the interference problem. RSMA provides precisely such a layer, as its rate-splitting structure is inherently efficient and flexible for multi-user interference management \cite{ref6, ref9}. When combined with FAS, this interference-aware design allows the system to translate port reconfiguration gains into tangible reductions in multi-user interference and substantial improvements in SE \cite{ref15}. Consequently, RSMA enables FAS to operate not only as a source of spatial diversity but also as a systematically exploited resource for interference-limited multi-user networks. In the Table~\ref{tab:FAS_RSMA_comparison}, FAS-RSMA is thus characterized by \emph{very strong} interference management ability, in contrast to the FAS-NOMA.

\subsubsection{\textbf{Relaxing CSI requirements}}
Accurate instantaneous CSI for all potential FAS antenna positions is difficult to obtain, especially when the number of candidate ports is large, the coherence time is short, or the system operates in high-mobility scenarios. In such conditions, the overhead and estimation errors associated with fully exploiting FAS can become prohibitive. RSMA is inherently robust to imperfect CSIT, since it can tolerate residual interference by flexibly adjusting the message split and power allocation \cite{ref6,ref10}. Thus, incorporating RSMA enhances the robustness of FAS-based systems and alleviates the CSI accuracy requirements at the BS \cite{ref11}. This robustness is reflected by the \emph{Very high} robustness level of FAS-RSMA in Table~\ref{tab:FAS_RSMA_comparison}, whereas NOMA-based schemes remain more sensitive to CSIT errors.

\subsubsection{\textbf{Enabling multi-objective designs}}
FAS naturally supports port reconfiguration diversity, but exploiting this diversity in a way that systematically accounts for fairness, security\footnote{{The common stream is decodable by all scheduled users, so it should not carry confidential payload when users are untrusted. In practice, confidentiality can be ensured via higher-layer encryption, while physical-layer secrecy can be enforced through secrecy-aware private-stream design \cite{ref1}.}}, or sensing constraints requires a higher-layer resource-allocation mechanism \cite{ref6}. RSMA provides such a mechanism by allowing the common stream to be used as a system-wide control lever. RSMA also enriches the design space of FAS-enabled secure and sensing-aware systems \cite{ref10}. In FAS-aided RSMA-ISAC architectures, the common stream can be exploited as a dual-purpose waveform that simultaneously conveys useful data to legitimate users and performs radar sensing functions \cite{ref1,ref10}. This multi-objective capability further differentiates FAS-RSMA from FAS-NOMA, which lack an equally flexible control handle at the signaling level.

Overall, FAS and RSMA play complementary roles: FAS provides hardware-level spatial DoFs via port reconfiguration, while RSMA exploits these DoFs at the signaling level to manage interference and diverse objectives. As Table~\ref{tab:FAS_RSMA_comparison} shows, this synergy uniquely achieves dynamic adaptation, robust interference management, and high diversity.

\section{Applications and Challenges of FAS-RSMA}
This section surveys representative application domains where FAS-RSMA can provide tangible performance gains over existing multiple-access schemes, and then identifies key open challenges that must be addressed before FAS-RSMA can be deployed at scale.
\subsection{Potential Applications of FAS-RSMA Systems}
\begin{figure*}
    \setlength{\belowcaptionskip}{-0.9cm}
    \centering
    \includegraphics[width=1.0\textwidth]{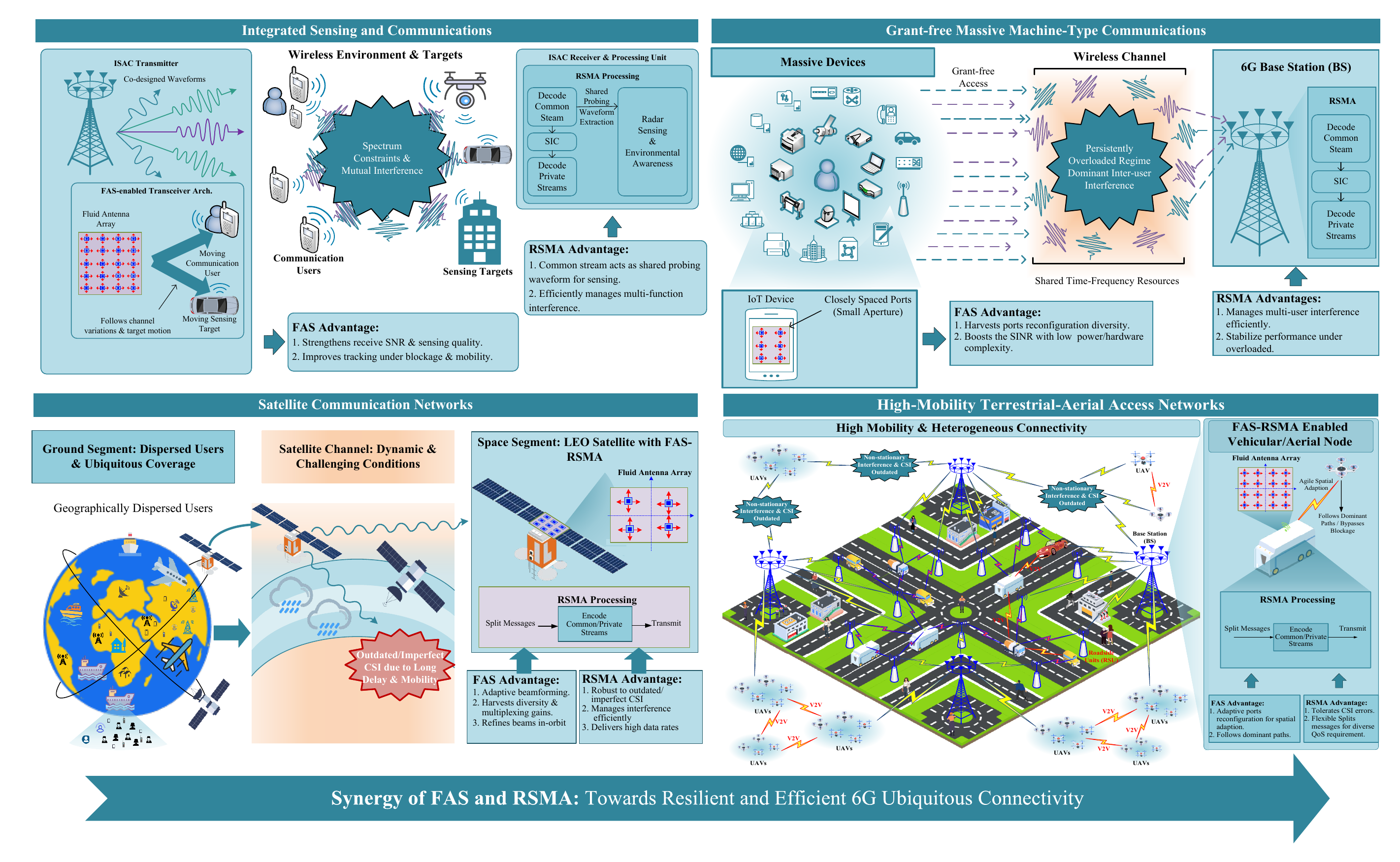}
    \caption{Potential Applications of FAS-RSMA Systems.}
    \label{fig:poential}
\end{figure*}

In this section, we outline several scenarios in which FAS-RSMA is particularly attractive, as illustrated in Fig.~\ref{fig:poential}.

\subsubsection{\textbf{Integrated sensing and communications (ISAC)}}
ISAC has emerged as a key paradigm for 6G, aiming to jointly provide high-quality connectivity and robust sensing capability. By co-designing waveforms and signal processing for both functions, ISAC consolidates communication and radar within a unified system and thereby improves spectral and hardware efficiency. Nevertheless, the concurrent operation of both functions places stringent constraints on spectrum and resource management and makes interference control more challenging than in communication-only deployments.

FAS-RSMA offers a promising way to relieve these tensions. On the FAS side, adaptive port reconfiguration enables the transceiver to reshape its effective aperture and follow channel variations and target motion, strengthening both receive SNR and sensing quality. On the RSMA side, the common stream decoded by many terminals not only manages inter-user interference, but can also be reused as a shared probing waveform for environmental awareness and radar sensing. In this way, FAS-RSMA can boost the reliability of both communication and sensing by coupling port diversity with interference-aware rate-splitting, leading to more accurate sensing and more robust links.

\subsubsection{\textbf{Grant-free massive machine-type communications}}
In 6G networks, massive machine-type communications (mMTC) will provide connectivity to tens of billions of devices whose traffic is low-power and sporadic, yet subject to diverse QoS targets. In such settings, arrivals are predominantly short packets generated by numerous uncoordinated terminals. Hence, grant-free access is indispensable to curb control signaling, enable concurrent reliable access, and limit contention. The sharing of common time-frequency resources by many devices drives the system into a persistently overloaded regime, in which inter-user interference becomes the dominant impairment. 

FAS-RSMA offers a natural way to tackle these challenges. On the one hand, RSMA is well-suited to overloaded operation: by embedding a common stream through which part of the interference is decoded, it can stabilize performance under heavy contention and diverse QoS constraints. Complementarily, FAS exposes many closely spaced candidate ports over a small aperture. By selecting the most favorable port per coherence interval, the device harvests reconfiguration diversity against the fading and impairments that are particularly harmful to short-packet links. Since port switching entails low power and low complexity, FAS meets the hardware budget requirement of mMTC while boosting SINR. Taken together, FAS-RSMA provides a practical pathway to grant-free mMTC under overload conditions: rate splitting mitigates multi-user interference, and port reconfiguration improves link quality without requiring additional RF chains.
\subsubsection{\textbf{Satellite communication networks}}
Satellite communication networks are expected to play a pivotal role in 6G by delivering ubiquitous coverage and service continuity in remote or infrastructure-poor regions. In particular, low Earth orbit (LEO) constellations are characterized by high satellite mobility, large propagation delays, and geographically dispersed users, leading to outdated and imperfect CSIT. Signals also experience attenuation due to atmospheric gases, rain, and cloud coverage. These dynamic and diverse conditions call for adaptive and robust interference-management mechanisms that can preserve link reliability and SE.

The FAS-RSMA paradigm is well aligned to cope with these issues. {To explicitly address the main bottlenecks, the performance gains are distinctly attributed: RSMA provides resilience against outdated CSIT, whereas FAS counters physical signal attenuation.} On the RSMA side, RSMA can maintain high data rates for multiple users even under imperfect CSIT, thereby mitigating performance losses caused by rapid channel evolution and delayed feedback in LEO constellations. On the FAS side, FAS-based reconfigurable apertures on satellites enable in-orbit or on-ground adaptation of effective element positions to refine beam patterns and harvest diversity and multiplexing gains. By unifying RSMA’s interference resilience with FAS’s spatial adaptability, FAS-RSMA can improve SE and robustness for satellite communications while keeping antenna hardware and signaling overhead within practical bounds.

\subsubsection{\textbf{High-Mobility Terrestrial-Aerial Access Networks}}
High-mobility access in 6G will involve heterogeneous devices on both the ground (vehicular) and in the air (UAV). Despite their different platforms, these systems share common characteristics: rapidly time-varying channels, pronounced Doppler shifts, and intermittent blockage by buildings or obstacles. Under such dynamics, conventional multi-user beamforming and scheduling struggle to maintain accurate CSIT, stabilize inter-cell and inter-user interference, and guarantee reliable links for diverse QoS requirements.

FAS-RSMA is well positioned to address these challenges in a unified manner. FAS arrays deployed at BS and/or on vehicular and aerial nodes can reconfigure their active antenna positions to follow dominant propagation paths, circumvent blockage, and adapt to changes in elevation, providing agile spatial adaptation without increasing the number of RF chains. On top of this, RSMA offers a robust downlink multiple-access strategy that tolerates imperfect CSIT and fluctuating interference through flexible splitting of messages into common and private streams. In combination, FAS-RSMA turns port-level spatial reconfiguration and interference-aware rate-splitting into complementary tools for sustaining reliable, fair, and interference-limited connectivity in integrated terrestrial–aerial high-mobility 6G networks.

\subsection{Major Research Challenges and Opportunities}
{The previous sections have shown that FAS and RSMA bring mutual benefits and can support a broad range of 6G applications.} However, FAS-RSMA technology is still in its infancy, and several open problems need to be solved before it can be adopted in practice.
\subsubsection{\textbf{Design of beamforming and FA location}}
The FAS-RSMA transmitter must jointly determine the common-private power and rate split, multi-user precoders, and the set of active antenna positions. This enlarged decision space, compounded by rapid channel variations and imperfect CSIT, undermines the scalability of classical RSMA solvers. Introducing FAS adds discrete DoFs that must be coordinated with continuous variables (beamforming, power, and rate), producing a high-dimensional, mixed-integer, strongly coupled, and nonconvex optimization problem. The joint design of beamforming and port reconfiguration is computationally intractable and remains difficult to scale.

Tractable solutions will likely rely on some ideas. First, a two-timescale control framework in which port reconfiguration follows a slow statistical policy while beamforming and rate are updated every slot. Second, simplify the search with structure-aware relaxations that encourage only a small set of antenna positions to be active, i.e., by adding soft penalties that nudge the solution toward sparsity. {Third, apply learning-based controllers tailored to this two-timescale structure to handle the nonconvex and stochastic nature of the problem. Specifically, deep reinforcement learning is promising for optimizing the slow-timescale discrete port-selection policy based on long-term channel statistics. Meanwhile, model-driven deep learning based on algorithm unrolling is promising for the fast-timescale continuous beamforming layer, since it can approximate iterative optimization algorithms with low online complexity and inference delay.}

\subsubsection{\textbf{Channel estimation}}
{Accurate acquisition of CSI is fundamental to realizing the gains of FAS-RSMA. However, the high mobility and frequent port reconfiguration inherent to FAS introduce significant challenges, particularly exacerbating feedback delay and outdated CSIT. Furthermore, a port-dense aperture enlarges the sounding space, which not only introduces spatial estimation errors but leads to overhead-limited sounding. This overhead limitation acts as the dominant bottleneck in FAS, severely driving up pilot overhead, feedback, and computational load, thereby creating a critical scalability issue since these resource requirements scale linearly with the growing number of devices and candidate ports. The RSMA structure adds another layer of difficulty: since user data is split into common and private parts, stream-aware CSI is required to capture cross-user interference and the coupling between the common stream and per-port channels, which in turn demands more intricate signaling and processing.}

{To aggressively address this massive overhead and avoid exhaustive estimation, promising directions include exploiting spatial correlation and port clustering to reduce the effective sounding dimension (e.g., probing only a subset of representative antenna positions and interpolating the rest), as well as compressive or on-off training schemes that first identify a small set of strong antenna positions and then refine their CSI. In high-mobility regimes, predictive CSI tracking based on filtering or learning-aided models can further reduce pilot overhead by leveraging temporal structure. Overall, designing low-overhead, low-latency CSI acquisition and management schemes that are explicitly tailored to FA dynamics and RSMA’s common and private structure remains an important and promising avenue in FAS-RSMA research.}

\subsubsection{\textbf{FAS-RSMA hardware design}}
The hardware design of FAS-RSMA poses several intertwined challenges. First, deciding how many FAS antenna positions to deploy, how to arrange them within a constrained aperture, and how many RF chains to allocate is non-trivial: enlarging the port set increases selection diversity and reliability, but also raises hardware cost and spatial correlation between antenna positions. Second, the use of RSMA with SIC places stricter requirements on the receiver front-end. Even with a single SIC stage, performance becomes more sensitive to analog non-idealities (e.g., nonlinearities, IQ imbalance, phase noise) and finite-resolution quantization, which can trigger error propagation. 

These challenges suggest several representative avenues for FAS-RSMA hardware design. For port deployment and RF-chain budgeting, correlation-aware design rules and simple parametric models can be used to choose a moderate number of antenna positions and their spacing, achieving most of the selection-diversity gain without excessive mutual coupling or cost. To make RSMA more tolerant to hardware imperfections, impairment-aware precoding and power-splitting designs, together with error-aware receivers and robust SIC algorithms, can explicitly account for front-end distortion and quantization effects.

\section{Case study}
In this section, we provide numerical case studies to quantify the performance gains of FAS-RSMA over its benchmarks under different antenna configurations. Specifically, we consider multi-user SISO and multi-user MISO systems, where a BS equipped with a single or multiple FPA serves multiple users. The user-side antenna array adopts a one-dimensional linear FAS architecture with maximum channel gain port reconfiguration. {The spatial correlation among the candidate ports is modeled by the Jakes autocorrelation model, which yields a Toeplitz covariance matrix on the one-dimensional linear aperture [15].} 

\subsection{Multi-user SISO systems}
We first consider a downlink multi-user SISO configuration. The SISO FAS-RSMA system is compared against FPA-RSMA, FAS-NOMA, and FPA-NOMA baselines to isolate the individual and joint contributions of FAS and RSMA. 

\begin{figure}[!t]
\setlength{\belowcaptionskip}{-0.9cm} %
\vskip 0.1in
\begin{center}
\centerline{\includegraphics[width=0.43\textwidth]{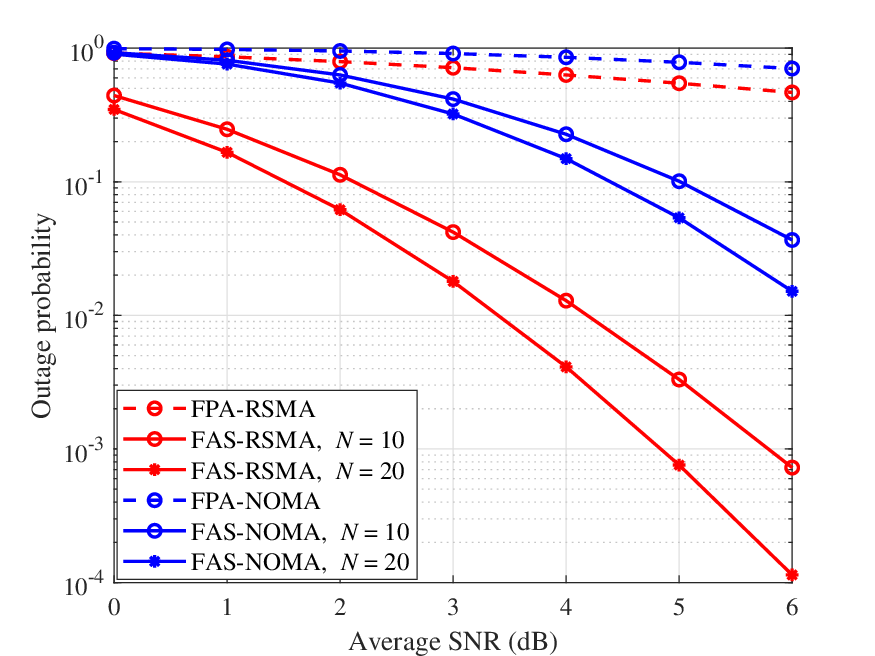}}
\caption{{OP comparison for $K=3$ users SISO downlink FAS-RSMA system and benchmark schemes, where $N$ is the number of candidate ports and the target rate thresholds follow \cite{ref15}.}}
\vspace{-10pt}
\label{fig:SISO_OP}
\end{center}
\vskip -0.1in
\end{figure}

Fig.~\ref{fig:SISO_OP} depicts the OP performance as a function of the average SNR. Several trends are apparent. First, both FPA-based baselines (FPA-RSMA and FPA-NOMA) exhibit relatively slow OP decay with SNR and remain in a high-OP regime over the entire range. For the single-antenna FPA configuration, this behavior reflects the limited diversity available without FAS, so neither RSMA nor NOMA can meet the stringent OP target across the full SNR range. Second, introducing FAS already brings noticeable gains for NOMA: FAS-NOMA with $N=10$ and $N=20$ achieves lower OP than FPA-NOMA, and the improvement is modest, as power-domain interference remains the main performance limiter. In contrast, FAS-RSMA achieves orders-of-magnitude OP reduction compared with all benchmarks, and its performance improves markedly as the number of FAS antenna positions increases (e.g., from $N=10$ to $N=20$), revealing that port reconfiguration diversity is effectively converted into reliability gains through rate-splitting. 

Overall, Fig.~\ref{fig:SISO_OP} confirms that, even in the multi-user SISO setting, the joint use of FAS and RSMA provides a much stronger reliability enhancement than using either FAS (FAS-NOMA) or RSMA (FPA-RSMA) in isolation. For systems targeting stringent reliability under practical hardware and power constraints, FAS-RSMA with a sufficiently large set of switchable antenna positions appears to be a particularly attractive design option.

\subsection{Multi-user MISO systems}
\begin{figure}[ht]
\setlength{\belowcaptionskip}{-0.9cm} %
\vskip 0.1in
\begin{center}
\centerline{\includegraphics[width=0.43\textwidth]{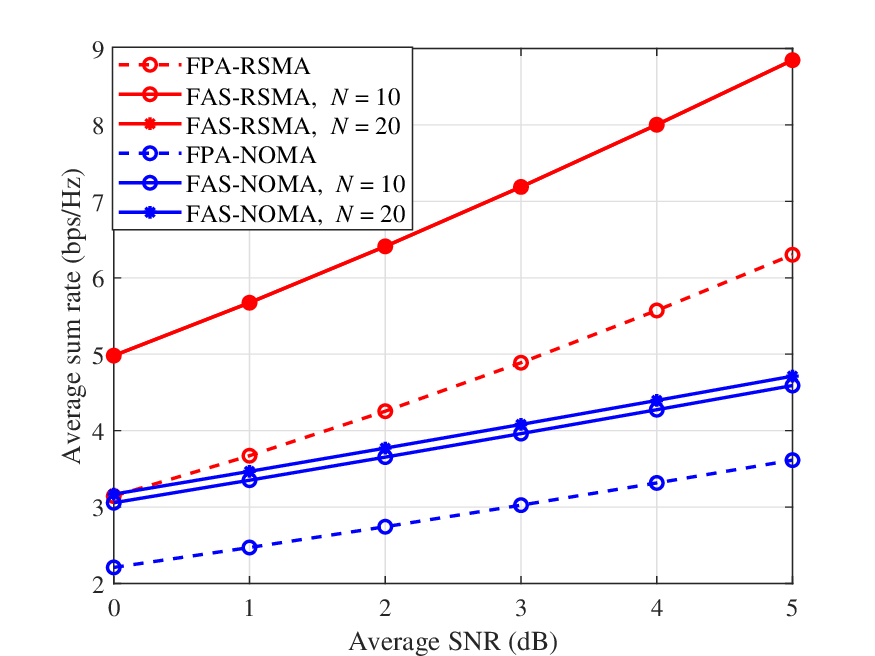}}
\caption{{Average sum rate comparison for $K=3$ users MISO downlink FAS-RSMA system and benchmark schemes with $L=4$ transmitted antennas, where \(N\) is the number of candidate ports.}}
\label{fig:MISO_AR_users}
\end{center}
\vskip -0.1in
\end{figure}

{We next consider a three-user MISO downlink configuration in which the BS employs linear precoding following the hybrid zero-forcing and maximum-ratio transmission principle, which serves as a practical and representative benchmark for RSMA-enabled multi-user MISO transmission.} We then compare the proposed FAS-RSMA scheme against the FPA-RSMA, FAS-NOMA, and FPA-NOMA baselines. The average sum rate performance is reported in Fig.~\ref{fig:MISO_AR_users}.

Several observations are in order. First, for both FPA- and FAS-based implementations, RSMA consistently achieves a higher sum rate than NOMA, confirming the SE advantage of rate-splitting over NOMA under the same RF-chain budget. Second, introducing FAS yields a clear rate offset for both access strategies, but the gain is markedly more pronounced for RSMA: the FAS-RSMA curves (red solid lines for $N=10$ and $N=20$) lie well above their FPA-RSMA counterpart (red dashed line) and all NOMA-based curves across the entire SNR range.

These results highlight a strong synergy between FAS and RSMA in the MISO setting. By dynamically reconfiguring the antenna position within the fluid-antenna aperture, FAS improves the effective channel gains and mitigates deep fades relative to FPA, while RSMA efficiently converts these improved channel conditions into SE gains through flexible interference management. FAS-RSMA therefore offers the highest average sum rate among the considered schemes.

Both multi-user SISO and MISO cases confirm the qualitative insights developed in the previous sections: FAS supplies low-cost spatial DoFs through port reconfiguration, while RSMA turns these DoFs into robust interference management and diversity gains. The results demonstrate that FAS-RSMA can substantially improve reliability and SE over FPA-RSMA and NOMA-type benchmarks, even under the linear port reconfiguration and precoding strategies, thereby supporting its promise as a practical candidate for 6G deployments.

\section{Conclusion}
This paper examined the emerging synergy between FAS and RSMA and showed how their complementary strengths can address key physical-layer challenges in 6G. FAS improves spatial diversity and strengthens weak links through low-cost port reconfiguration, while RSMA provides robust interference management under imperfect CSIT and heterogeneous user conditions. Through a structured review, application outlook, and case studies, we highlighted the potential of FAS-RSMA in terms of reliability, throughput, and adaptability. Important open issues remain in joint beamforming and FA-location design, channel estimation, and hardware implementation. Addressing these challenges will be essential to translating FAS-RSMA into practical 6G deployments.

\newpage

 




\vfill

\end{document}